\documentclass[letterpaper]{article} 
\usepackage{aaai2027}  
\usepackage[hyphens]{url}  
\usepackage{graphicx} 
\usepackage{natbib}  
\usepackage{caption} 
\usepackage{algorithm}
\usepackage{algorithmic}

\usepackage{newfloat}
\usepackage{listings}
\DeclareCaptionStyle{ruled}{labelfont=normalfont,labelsep=colon,strut=off} 
\floatstyle{ruled}
\newfloat{listing}{tb}{lst}{}
\floatname{listing}{Listing}

\usepackage{booktabs}

\newcommand{\method}{HyperAgent4POI}
\usepackage{amsmath}
\usepackage{amssymb}
\usepackage{multirow}
\usepackage{graphicx}
\usepackage{subcaption}

\title{HyperAgent4POI: Dynamic Semantic Message Passing on Multi-Agent Hypergraphs for Missing-Modality Recommendation}
\author {
    Jinze Wang\textsuperscript{\rm 1,2}\equalcontrib,
    Yuze Liu\textsuperscript{\rm 1}\equalcontrib,
    Tiehua Zhang\textsuperscript{\rm 2}\corresponding
    Jiong Jin\textsuperscript{\rm 1},
    Zhu Sun\textsuperscript{\rm 3}
}
\affiliations {
    \textsuperscript{\rm 1}Swinburne University of Technology, Melbourne, Australia\\
    \textsuperscript{\rm 2}Tongji University, Shanghai, China\\
    \textsuperscript{\rm 3}Singapore University of Technology and Design, Singapore\\
}

\begin{document}

\maketitle

\begin{abstract}
Next Point-of-Interest (POI) recommendation benefits from textual and visual content that describes venue semantics, yet such content is often incomplete in real-world services. Missing modalities weaken POI representations and reduce the semantic evidence available for ranking. The resulting representations also provide unreliable evidence for modeling higher-order user--POI interactions. We propose HyperAgent4POI, which uses Dynamic Semantic Message Passing (DSMP) to perform modality completion and soft incidence refinement within each hypergraph layer. Persistent node agents share a frozen Llama backbone and use role-specific adapters to produce node-to-hyperedge messages. Semantic hyperedge motifs formed from these messages guide soft incidence scoring and modality completion. Final node representations are cached for online ranking without LLM calls. Experiments on three real-world LBSN datasets show consistent ranking gains over 15 baselines across modality-missing rates, while cached inference provides practical online efficiency. Under a 60\% modality-missing rate, HyperAgent4POI improves NDCG@20 over the strongest baseline by 8.2\% on average across the three datasets.
\end{abstract}


\section{Introduction}
\label{sec:intro}

Location-Based Social Networks (LBSNs) record user check-ins at Points-of-Interests (POIs), such as restaurants, museums, parks, and transit hubs~\cite{wang2023meta,wang2025hyperman}. Effective next POI recommendation must model both individual user--POI interactions and high-order dependencies among temporally related visits. A user's visits within a short period often form a coherent activity pattern. Pairwise graphs represent each interaction as one edge and require additional construction to encode such co-visit groups~\cite{yan2023spatio}. Hypergraphs provide a more direct representation by connecting a user and multiple temporally related POIs with one hyperedge. They have therefore become useful for POI and collaborative recommendation~\cite{lv2021we,xia2022hypergraph,ji2020dual}.

Beyond interaction structure, next POI recommendation relies on POI content to distinguish venues and match them to user preferences. Real venues have uneven coverage of text and image modalities, while stable attributes such as geography and category are usually easier to observe~\cite{chen2026modalities,kim2025disentangling}. When text or images are unavailable, the model must rank candidates from incomplete semantic evidence, and common substitutes such as placeholders, zeros, or mean features distort the initial POI representation. Prior graph-learning studies report that standard GNN formulations assume feature availability and that separating imputation from graph learning can degrade performance~\cite{taguchi2021graph,rossi2022unreasonable}. This problem becomes more consequential in hypergraph message passing: the representation discrepancy enters the first hypergraph layer, and node-to-hyperedge and hyperedge-to-node aggregation repeatedly transmit it to neighboring nodes and subsequent soft incidence updates. We refer to this layer-wise transmission as \emph{discrepancy propagation}.

Existing approaches address different parts of this problem. Several modality-completion methods reconstruct features before or outside graph propagation~\cite{li2025generating,zhao2025graph,ding2025learnable}, so their completed features are fixed before the model observes a layer-specific hyperedge context. Dynamic hypergraph methods refine incidences from learned embeddings or interaction geometry~\cite{lin2025unified,chen2020hgmf}, but their topology scores may depend on the same incomplete representations used for prediction. Recommenders augmented with large language models (LLMs) enrich item or user representations~\cite{pipoli2025missrag,kuai2026generating}, yet feature-level augmentation outside message passing does not couple modality completion with soft incidence refinement. Thus, the two operations are not jointly conditioned on the current layer-specific hyperedge context.

To address these gaps, we propose HyperAgent4POI, an agent-based dynamic semantic message-passing framework. Each graph node is represented by a persistent agent that carries its state across DSMP layers. Within every layer, DSMP applies four role-conditioned semantic operators to the agent states and current hyperedge context. Each operator combines a shared frozen Llama backbone with a trainable role-specific LoRA adapter and structured role inputs. All agents therefore share model parameters while maintaining distinct node states. Local Expert transforms each node-agent state into a node-to-hyperedge message, while Semantic Aggregator summarizes these messages into a semantic hyperedge motif. They use different role inputs and share the aggregation adapter. Topology Evolution and Modality Deduction use separate adapters to refine bounded soft incidences and complete unavailable modality features, respectively. The completed features update node-agent states for the next layer. Together, these operations define DSMP as a layer-wise loop for semantic aggregation, soft incidence refinement, and modality completion.

DSMP differs from one-shot completion in when hyperedge context is used. Pre-propagation completion produces a fixed feature before message passing, whereas DSMP revisits each unavailable modality after soft incidence refinement and supplies the completed feature to the next layer. The LLM operators run only during training and offline representation refresh. Final node representations are cached for dot-product ranking, so online next POI serving requires no LLM calls.

Our contributions are:
\begin{enumerate}
  \item We formalize incomplete multimodal next POI recommendation on interaction hypergraphs. A motivating linearized analysis illustrates how representation discrepancies affect hypergraph propagation.
  \item We propose DSMP, which uses role-conditioned LLM hidden states to construct node-to-hyperedge messages and semantic hyperedge motifs, refine bounded POI--hyperedge incidences, and complete unavailable modalities within each hypergraph layer.
  \item Experiments on three LBSN benchmarks show consistent gains across modality-missing rates. Under a 60\% modality-missing rate, HyperAgent4POI improves NDCG@20 over the strongest baseline by 8.2\% on average across the three datasets.
\end{enumerate}


\section{Related Work}
\label{sec:related}

\textbf{Incomplete Multimodal Recommendation.} Multimodal recommenders combine collaborative and content signals, but missing modalities leave uneven evidence across POIs~\cite{wang2026meta}. Earlier methods use modality dropout, autoencoding, or shared spaces~\cite{wang2018lrmm,ganhor2024multimodal}; recent work reconstructs missing representations explicitly. DGMRec separates shared and modality-specific factors~\cite{kim2025disentangling}, MoDiCF combines diffusion reconstruction with counterfactual learning~\cite{li2025generating}, and graph retrieval supplies nonlocal completion context~\cite{li2026robust}. HIRE already couples completion with heterogeneous-hypergraph construction and sparsification~\cite{lin2025unified}. DSMP differs in how this coupling is implemented: role-conditioned LLM hidden states operate on the current node--hyperedge context to update node-to-hyperedge messages, soft incidences, and missing-modality states within each layer.

\noindent\textbf{Hypergraph Next POI Recommendation.} DHCF and HCCF model high-order collaborative dependencies~\cite{ji2020dual,xia2022hypergraph}, while DHLCF learns differentiable incidences regularized by the interaction graph~\cite{wei2022dynamic}. Task-specific models add mobility structure: MSTHN combines local spatial--temporal graphs with a global hypergraph~\cite{lai2023multi}; ReHDM constructs region-aware trajectory hyperedges~\cite{li2025beyond}; MSAHG learns scenario-specific multi-view sub-hypergraphs~\cite{lin2026multifaceted}; and HGDRec uses global hypergraphs with diffusion-optimized trajectory intent~\cite{pan2025hgdrec}. HyperSE fuses pretrained-language-model semantic features with global and local POI hypergraphs~\cite{zeng2025global}, but its semantic enhancement is external to a recursive missing-modality update. DSMP instead conditions incidence scoring and completion on the semantic hyperedge motif produced at the current propagation layer.

\noindent\textbf{LLM-Augmented Graph and POI Recommendation.} Feature-level methods prompt LLMs to enrich item descriptions~\cite{lyu2024llm} or align generated profiles with collaborative representations~\cite{ren2024representation}. Graph-aware methods augment interactions and contrastive supervision~\cite{zheng2025lagcl4rec}, encode graph-constructed high-order interactions~\cite{wang2024enhancing}, or express multimodal graphs as graph-of-thought prompts~\cite{yi2025multi}. LLMHG combines LLM preference semantics with multi-view hypergraph learning~\cite{chu2024llm}, while HeLLM injects multimodal hypergraph representations into an LLM~\cite{guo2025multi}; agentic recommenders assign roles for profiling, retrieval, reflection, and ranking~\cite{wang2024macrec,peng2025survey,wang2025we}. In POI recommendation, ZeroPOIRec performs zero-shot LLM ranking~\cite{kim2025large}, Refine-POI reinforcement-fine-tunes an LLM for next POI ranking~\cite{li2025refine}, and Agent4POI generates query-conditioned multimodal affordance representations~\cite{wang2026agent4poi}. These methods enrich inputs, optimize direct ranking, or construct query-time item representations. HyperAgent4POI uses frozen-LLM hidden states as role-conditioned semantic operators with fixed message interfaces inside each hypergraph layer.

\section{Methodology}
\label{sec:method}

\begin{figure*}[t]
\centering
\includegraphics[width=0.95\textwidth]{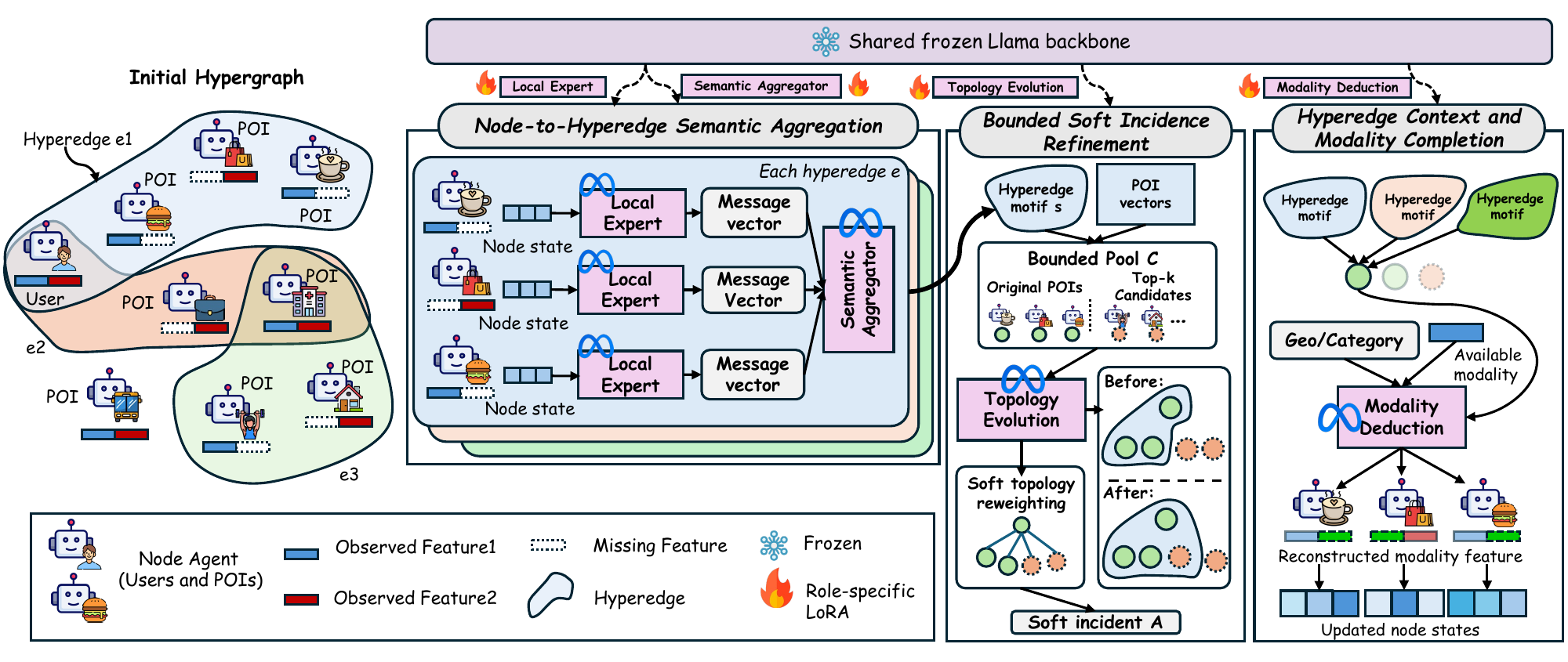}
\caption{Overall architecture of HyperAgent4POI and its layer-wise DSMP workflow. Each DSMP layer uses role-conditioned LLM hidden states to aggregate node evidence into semantic hyperedge motifs, refine bounded POI–hyperedge incidences, and complete unavailable modalities. The completed features and refined hyperedge context are then used to update node states for the next layer.}
\label{fig:framework}
\end{figure*}

\subsection{Problem Formulation}
\label{sec:problem}
Let \(\mathcal{U}\) and \(\mathcal{P}\) be the user and POI sets, and let \(\mathcal{V}=\mathcal{U}\cup\mathcal{P}\). We construct an initial hypergraph \(\mathcal{G}^{(0)}=(\mathcal{V},\mathcal{E},\mathbf{H}^{(0)})\) from check-in sessions. Each hyperedge \(e\in\mathcal{E}\) contains its anchor user \(u_e\) and the POIs visited in that session; \(\mathbf{H}^{(0)}\in\{0,1\}^{|\mathcal{V}|\times|\mathcal{E}|}\), where \(\mathbf{H}_{i,e}^{(0)}=1\) if and only if node \(i\) belongs to \(e\). Each POI \(p\) has geographical and categorical attributes \((x_p^{\mathrm{geo}},x_p^{\mathrm{cat}})\) and semantic modalities indexed by \(\mathcal{K}\). Let \(\mathcal{S}\subseteq\mathcal{P}\times\mathcal{K}\) contain the POI--modality pairs for which source content exists, and let \(\mathcal{O}\subseteq\mathcal{S}\) contain the pairs exposed to the recommender. For \((p,k)\in\mathcal{S}\), a frozen encoder produces \(\mathbf{z}_p^k=f_{\mathrm{enc}}^k(y_p^k)\in\mathbb{R}^{d_k}\), where \(y_p^k\) is the raw modality content and \(d_k\) is its encoded dimension. The unavailable pairs are \(\mathcal{R}=(\mathcal{P}\times\mathcal{K})\setminus\mathcal{O}\), while \(\mathcal{M}=\mathcal{S}\setminus\mathcal{O}\) contains masked pairs that provide completion targets during training. Naturally unavailable pairs belong to \(\mathcal{R}\setminus\mathcal{M}\) and never provide target features.

Given a training triple \((u,p^+,p^-)\in\mathcal{U}\times\mathcal{P}\times\mathcal{P}\), \(p^+\) is the chronologically next POI in a training-prefix sequence and \(p^-\) is unvisited by \(u\). The primary task ranks \(p^+\) above \(p^-\). Validation and test rank each held-out next interaction from the fixed history available at the training cutoff; the hypergraph and cached node representations remain unchanged. The auxiliary task reconstructs \(\mathbf{z}_p^k\) only for \((p,k)\in\mathcal{M}\), coupling ranking with supervised completion without exposing masked targets.

\paragraph{Discrepancy propagation.}
A missing-modality embedding gives each unavailable POI--modality pair a defined input but generally differs from its complete-input representation. To trace how this discrepancy enters hypergraph propagation, consider a static, linear random-walk-normalized layer. If \(\boldsymbol{\Delta}^{V,(l)}\) is the difference between incomplete- and complete-input node representations at layer \(l\), then
\begin{equation}
\boldsymbol{\Delta}^{V,(l+1)}
=\underbrace{\mathbf{D}_V^{-1}\mathbf{H}^{(0)}\mathbf{W}_E\mathbf{D}_E^{-1}(\mathbf{H}^{(0)})^\top}_{\mathbf{P}_{\mathcal G}}
\boldsymbol{\Delta}^{V,(l)}\boldsymbol{\Theta}^{(l)} ,
\label{eq:error_propagation}
\end{equation}
where \(\mathbf{W}_E\) is the diagonal hyperedge-weight matrix, \([\mathbf{D}_V]_{ii}=\sum_e\mathbf{H}_{i,e}^{(0)}[\mathbf{W}_E]_{ee}\), \([\mathbf{D}_E]_{ee}=\sum_i\mathbf{H}_{i,e}^{(0)}\), and \(\boldsymbol{\Theta}^{(l)}\) is the feature transform. The normalized operator \(\mathbf{P}_{\mathcal G}\) isolates how an existing representation discrepancy is transmitted to neighboring nodes; DSMP generalizes this motivating linearization with nonlinear, layer-dependent semantic updates.

\subsection{Dynamic Semantic Message Passing}
\label{sec:dsmp}

Let \(L\) be the number of DSMP layers, indexed by \(l\in\{0,\ldots,L-1\}\), and let \(\varepsilon_{\mathrm{num}}>0\) be a numerical stabilizer. HyperAgent4POI places semantic aggregation, soft incidence refinement, and modality completion inside every hypergraph layer. Each node \(i\in\mathcal{V}\) has a persistent node agent \(a_i\) with state \(\mathbf{x}_i^{(l)}\). A node agent is a stateful semantic message processor: it maintains node-local state across layers, uses the shared Llama backbone with role-specific LoRA adapters~\cite{hu2022lora}, and communicates through the message interfaces below.

\begin{figure}[t]
\centering
\includegraphics[width=.88\columnwidth]{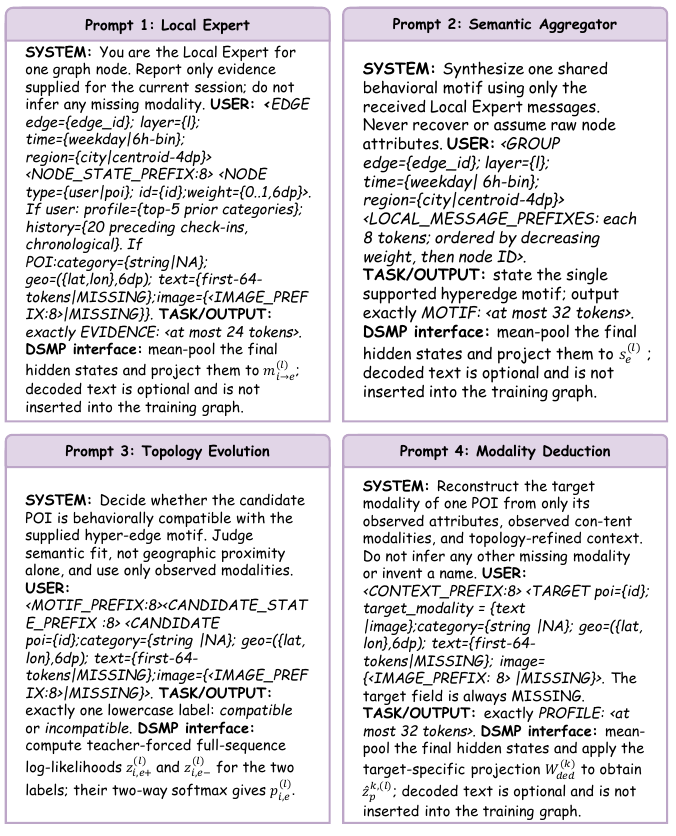}
\caption{Prompt templates for the four role-conditioned semantic operators in DSMP. Bracketed vectors denote continuous prefixes.}
\label{fig:prompt_template}
\end{figure}

The semantic operators share a frozen Llama-3 backbone with role-specific LoRA adapters, \(r\in\{\mathrm{agg},\mathrm{evo},\mathrm{ded}\}\). Local Expert and Semantic Aggregator use different inputs but share the aggregation adapter. For a discrete role input \(Q\) and optional continuous prefix \(\mathbf{P}\), \(\mathbf{h}_r(Q;\mathbf{P})\) denotes the mean-pooled final hidden state of adapter \(r\). Decoded text is used only for inspection; training and recommendation consume hidden states. Figure~\ref{fig:prompt_template} summarizes the four role templates. Prompt serialization and target-isolation details are provided in Supplementary Section~A. We write \(\operatorname{norm}(\mathbf{v})=\mathbf{v}/(\|\mathbf{v}\|_2+\varepsilon_{\mathrm{num}})\).

DSMP maintains soft incidences \(\mathbf{A}^{(l)}\in[0,1]^{|\mathcal{V}|\times|\mathcal{E}|}\), initialized by \(\mathbf{A}^{(0)}=\mathbf{H}^{(0)}\). Let \(\mathcal{V}_e^{(l)}\) be the bounded node group evaluated for hyperedge \(e\), with \(\mathcal{V}_e^{(0)}=\{i:\mathbf{H}_{i,e}^{(0)}=1\}\). For each modality, let \(\mathbf{z}_{\mathrm{MISS}}^k\) be a learned missing-modality embedding. The feature supplied before layer \(l\) is
\begin{equation}
\widetilde{\mathbf{z}}_p^{k,(l)}=
\begin{cases}
\mathbf{z}_p^k, &(p,k)\in\mathcal{O},\\
\mathbf{z}_{\mathrm{MISS}}^k, &(p,k)\in\mathcal{R},\ l=0,\\
\widehat{\mathbf{z}}_p^{k,(l-1)}, &(p,k)\in\mathcal{R},\ l>0.
\end{cases}
\label{eq:modality_state}
\end{equation}
Type-specific encoders initialize user states from user ID and history and POI states from ID, geographical and categorical attributes, and the features in Equation~\eqref{eq:modality_state}. Modality projections map these features to continuous LLM prefixes, so the backbone never receives raw non-text media. No role input contains \(y_p^k\) or \(\mathbf{z}_p^k\) for \((p,k)\in\mathcal{R}\).

\subsubsection{Node-to-Hyperedge Semantic Aggregation}
\label{sec:agg}

For each \(i\in\mathcal{V}_e^{(l)}\), the Local Expert maps the current node state, exposed attributes, incidence weight, and session context to an edge-specific message. Calls are independent across nodes; their outputs are ordered by decreasing incidence weight before aggregation:
\begin{equation}
\begin{aligned}
\mathbf{m}_{i\rightarrow e}^{(l)}
&=\operatorname{norm}\!\left(\mathbf{W}_{\mathrm{msg}}\mathbf{h}_{\mathrm{agg}}(Q_{i\rightarrow e}^{(l)};\mathbf{P}_i^{(l)})\right),\\
\mathbf{P}_e^{(l)}
&=\mathop{\Vert}_{i\in\mathcal{V}_e^{(l)}}\left[\mathbf{A}_{i,e}^{(l)}\mathbf{W}_{\mathrm{pre}}\mathbf{m}_{i\rightarrow e}^{(l)}\right],\\
\mathbf{s}_e^{(l)}
&=\operatorname{norm}\!\left(\mathbf{W}_{\mathrm{agg}}\mathbf{h}_{\mathrm{agg}}(Q_{e,\mathrm{agg}}^{(l)};\mathbf{P}_e^{(l)})\right).
\end{aligned}
\label{eq:agent_collaboration}
\end{equation}
Here, \(\mathbf{W}_{\mathrm{msg}}\), \(\mathbf{W}_{\mathrm{pre}}\), and \(\mathbf{W}_{\mathrm{agg}}\) are trainable linear maps; \(\mathbf{P}_i^{(l)}\) encodes \(\mathbf{x}_i^{(l)}\) and its exposed evidence; and \(\Vert\) denotes concatenation. The input \(Q_{e,\mathrm{agg}}^{(l)}\) contains session metadata, while node-specific evidence enters through \(\mathbf{P}_e^{(l)}\). The resulting \(\mathbf{s}_e^{(l)}\) is the semantic hyperedge motif passed from nodes to hyperedge \(e\).

\subsubsection{Bounded Soft Incidence Refinement}
\label{sec:evo}

Let \(\mathcal{N}_{e,\mathcal{P}}^{(0)}=\{p\in\mathcal{P}:\mathbf{H}_{p,e}^{(0)}=1\}\) be the original POI members, and let \(K_{\mathrm{ret}}\) be a positive integer retrieval budget. At layer \(l\), a frozen stop-gradient retriever returns the \(K_{\mathrm{ret}}\) POIs whose current profiles best match \(\mathbf{s}_e^{(l)}\). The candidate pool \(\mathcal{C}_e^{(l)}=\mathcal{N}_{e,\mathcal{P}}^{(0)}\cup\operatorname{TopK}_{K_{\mathrm{ret}}}(\mathbf{s}_e^{(l)})\) is recomputed from the original members at every layer. Hence, \(|\mathcal{C}_e^{(l)}|\leq|\mathcal{N}_{e,\mathcal{P}}^{(0)}|+K_{\mathrm{ret}}\), and candidates introduced at one layer do not recursively expand later pools.

For each \(p\in\mathcal{C}_e^{(l)}\), Topology Evolution scores the two label strings \(y_+=\texttt{compatible}\) and \(y_-=\texttt{incompatible}\). Let \(z_{p,e,y}^{(l)}=\log P(y\mid Q_{p,e,\mathrm{evo}}^{(l)},\mathbf{s}_e^{(l)})\) be the teacher-forced sequence log-likelihood. Training retains differentiable soft incidences:
\begin{equation}
\begin{aligned}
\pi_{p,e}^{(l)}
&=\frac{\exp z_{p,e,+}^{(l)}}{\exp z_{p,e,+}^{(l)}+\exp z_{p,e,-}^{(l)}},\\
\mathbf{A}_{i,e}^{(l+1)}
&=\begin{cases}1, & i=u_e,\\ \pi_{i,e}^{(l)}, & i\in\mathcal{C}_e^{(l)},\\ 0, & \text{otherwise},\end{cases}
\qquad \mathcal{V}_e^{(l+1)}=\{u_e\}\cup\mathcal{C}_e^{(l)} .
\end{aligned}
\label{eq:soft_rewire}
\end{equation}
Training and offline representation refresh use soft incidences. Given an export threshold \(\tau\in(0,1)\), the final POI--hyperedge hard incidence \(\mathbf{H}_{p,e}^{\mathrm{hard}}=\mathbb{I}[\pi_{p,e}^{(L-1)}>\tau]\) for \(p\in\mathcal{C}_e^{(L-1)}\) can be exported for storage and interpretation, where \(\mathbb{I}[\cdot]\) denotes the indicator function. If no POI survives, the highest-scoring candidate is retained. Recommendation uses the cached node representations computed with soft incidences.

\subsubsection{Hyperedge Context and Modality Completion}
\label{sec:ded}

The refined incidences propagate semantic hyperedge motifs back to nodes. For every unavailable pair \((p,k)\in\mathcal{R}\), Modality Deduction combines the refined hyperedge context with the POI's geographical and categorical attributes and its other exposed modalities. A modality-specific head then maps the shared hidden state to the target feature space:
\begin{equation}
\begin{aligned}
\mathbf{c}_i^{(l)}
&=\frac{\sum_{e\in\mathcal{E}}\mathbf{A}_{i,e}^{(l+1)}\mathbf{s}_e^{(l)}}{\varepsilon_{\mathrm{num}}+\sum_{e\in\mathcal{E}}\mathbf{A}_{i,e}^{(l+1)}},\\
\mathbf{d}_p^{k,(l)}
&=\mathbf{h}_{\mathrm{ded}}(Q_{p,k,\mathrm{ded}}^{(l)};\mathbf{W}_c\mathbf{c}_p^{(l)}),\\
\widehat{\mathbf{z}}_p^{k,(l)}
&=\operatorname{norm}(\mathbf{W}_{\mathrm{ded}}^k\mathbf{d}_p^{k,(l)}),\quad (p,k)\in\mathcal{R},\\
\bar{\mathbf{z}}_p^{k,(l)}
&=\begin{cases}
\mathbf{z}_p^k, &(p,k)\in\mathcal{O},\\
\widehat{\mathbf{z}}_p^{k,(l)}, &(p,k)\in\mathcal{R},
\end{cases}\\
\mathbf{q}_p^{(l+1)}&=\mathop{\Vert}_{k\in\mathcal{K}}\bar{\mathbf{z}}_p^{k,(l)},\\
\mathbf{x}_p^{(l+1)}
&=f_P^{\mathrm{upd}}\!\left([\mathbf{x}_p^{(l)};x_p^{\mathrm{geo}};x_p^{\mathrm{cat}};\mathbf{q}_p^{(l+1)};\mathbf{c}_p^{(l)}]\right),\\
\mathbf{x}_u^{(l+1)}
&=f_U^{\mathrm{upd}}\!\left([\mathbf{x}_u^{(l)};\mathbf{c}_u^{(l)}]\right).
\end{aligned}
\label{eq:state_update}
\end{equation}
Here, \(\mathbf{W}_c\) is a trainable context projection, \(\mathbf{W}_{\mathrm{ded}}^k\) maps the deduction state to \(\mathbb{R}^{d_k}\), and \(f_P^{\mathrm{upd}}\) and \(f_U^{\mathrm{upd}}\) are type-specific update networks. Equation~\eqref{eq:modality_state} then supplies each completed feature to the next layer. Consequently, completion uses the refined hyperedge context, and the completed evidence can affect subsequent semantic messages and incidence scores.

\subsection{Joint Learning and Deployment}
\label{sec:loss}

Type-specific readout networks \(g_t\), \(t\in\{\mathrm{user},\mathrm{POI}\}\), produce \(\mathbf{v}_i=g_{\operatorname{type}(i)}(\mathbf{x}_i^{(L)})\), where \(\operatorname{type}(i)\) denotes the node type. The ranking score is \(\widehat y_{u,p}=\mathbf{v}_u^\top\mathbf{v}_p\). Let \(\mathcal{D}\) denote the set of training triples. The BPR objective~\cite{rendle2012bpr} is
\begin{equation}
\mathcal{L}_{\mathrm{BPR}}=-\frac{1}{|\mathcal{D}|}\sum_{(u,p^+,p^-)\in\mathcal{D}}\log\sigma\!\left(\widehat y_{u,p^+}-\widehat y_{u,p^-}\right).
\label{eq:bpr}
\end{equation}
where \(\sigma\) is the logistic sigmoid.

For a minibatch, let \(\mathcal{B}_k=\{p:(p,k)\in\mathcal{M}\}\) and \(\mathcal{K}_{\mathcal B}=\{k\in\mathcal{K}:|\mathcal{B}_k|\geq2\}\). With cosine similarity \(\operatorname{sim}\) and temperature \(T>0\), the modality-balanced completion loss is
\begin{equation}
\begin{aligned}
s_{ij}^k&=\operatorname{sim}(\widehat{\mathbf{z}}_i^{k,(L-1)},\mathbf{z}_j^k)/T,\\
\mathcal{L}_{\mathrm{NCE}}
&=-\frac{1}{|\mathcal{K}_{\mathcal B}|}\sum_{k\in\mathcal{K}_{\mathcal B}}
\frac{1}{|\mathcal{B}_k|}\sum_{i\in\mathcal{B}_k}\\
&\quad \log\frac{\exp(s_{ii}^k)}{\sum_{j\in\mathcal{B}_k}\exp(s_{ij}^k)} .
\end{aligned}
\label{eq:infonce}
\end{equation}
Minibatches include at least two supervised targets for each sampled modality, ensuring \(\mathcal{K}_{\mathcal B}\neq\varnothing\).

To regularize topology refinement, fix \(\varepsilon_{\mathrm{lab}}\in(0,\tfrac12)\) and define \(\mathcal{P}^{(l)}_{\mathrm{prop}}=\{(p,e):p\in\mathcal{C}_e^{(l)}\}\), \(q_{p,e}=(1-2\varepsilon_{\mathrm{lab}})\mathbf{H}_{p,e}^{(0)}+\varepsilon_{\mathrm{lab}}\), and \(\bar a_{p,e}^{(l+1)}=\operatorname{clip}(\mathbf{A}_{p,e}^{(l+1)},\varepsilon_{\mathrm{lab}},1-\varepsilon_{\mathrm{lab}})\). We average the finite Bernoulli divergence across layers and proposed pairs:
\begin{equation}
\begin{aligned}
\mathcal{L}_{\mathrm{top}}
&=\frac{1}{L}\sum_{l=0}^{L-1}\frac{1}{|\mathcal{P}^{(l)}_{\mathrm{prop}}|}
\sum_{(p,e)\in\mathcal{P}^{(l)}_{\mathrm{prop}}}\\
&\quad D_{\mathrm{KL}}\!\left(\operatorname{Ber}(\bar a_{p,e}^{(l+1)})
\middle\|\operatorname{Ber}(q_{p,e})\right).
\end{aligned}
\label{eq:topology_loss}
\end{equation}
Here, \(\operatorname{Ber}(q)\) is the Bernoulli distribution with parameter \(q\), and \(D_{\mathrm{KL}}\) is the Kullback--Leibler divergence. This conservative edit prior penalizes unsupported departures from the initial interaction graph. The remaining objectives still permit additions, removals, and reweighting when they improve the joint objective.

The total objective is
\begin{equation}
\mathcal{L}=\mathcal{L}_{\mathrm{BPR}}+\lambda\left(\mathcal{L}_{\mathrm{NCE}}+\gamma\mathcal{L}_{\mathrm{top}}\right).
\label{eq:loss_total}
\end{equation}
The coefficients \(\lambda>0\) and \(\gamma>0\) weight the auxiliary completion and topology terms. The backbone, retriever, and modality encoders remain frozen. LoRA adapters, modality projections, update and readout networks, missing-modality and ID embeddings, and prediction heads are trained jointly; gradients do not pass through candidate retrieval. DSMP role calls run during offline representation refresh with soft incidences. Final node representations are cached for dot-product recommendation, and the thresholded topology can be cached separately for storage and interpretation. Figure~\ref{fig:framework} provides an overview of HyperAgent4POI and DSMP workflow.


\section{Experiments}
\label{sec:exp}

We design the experiments to answer five research questions: (RQ1)~How does HyperAgent4POI compare with representative next POI recommendation baselines when POI modalities are missing? (RQ2)~How robust is \method{} under different levels of modality missingness? (RQ3)~How much does each DSMP component contribute to recommendation performance? (RQ4)~Does layer-wise completion provide benefits beyond one-shot pre-completion? (RQ5)~What are the offline and online costs of \method{}?

\subsection{Experimental Setup}
\label{sec:setup}

\subsubsection{Datasets.}

Following~\cite{wang2025hyperman,wang2026agent4poi}, we evaluate three LBSN benchmarks. \emph{Yelp-2018}~\cite{jendal2025yelp} contains reviews, business metadata, and photos. \emph{Foursquare-NYC} (FSQ-NYC) and \emph{Foursquare-TKY} (FSQ-TKY)~\cite{yang2016participatory} provide check-ins and coordinates augmented with matched Google Places images and Yelp reviews. Additional matching and validation details are provided in Supplementary Section~B. Accepted-match precision is 96.8\% (95\% CI: 95.1\%--98.1\%), with multimodal content coverage of 97.4\% for NYC and 96.1\% for TKY. After filtering, interactions are split chronologically into 70\% training, 10\% validation, and 20\% test. Hyperedges, prompts, retrieval indices, cached node representations, and POI content use only data available by the training cutoff; crossing sessions are truncated at that cutoff. Table~\ref{tab:data} summarizes the data.

More specifically, sessions are formed per user by grouping consecutive check-ins separated by at most 24 hours. Each session induces a user--POI hyperedge, providing higher-order co-visit context beyond pairwise user--POI interactions. To simulate modality missingness, we identify the POI--modality pairs with available text or image features. For each random seed and modality-missing rate $\rho_{\mathrm{miss}}\in\{0.2,0.4,0.6,0.8\}$, we uniformly sample without replacement the corresponding proportion of these pairs and mask only their model inputs. A POI may lose its text modality, image modality, or both, while its geographical and categorical attributes remain observed. The masked features are retained only as completion targets and excluded from prompts, retrieval features, and recommendation inputs. Modalities unavailable in the original data remain masked and are excluded from completion evaluation. This protocol is used in all robustness experiments.

\begin{table}[t]
\centering
\small
\setlength{\tabcolsep}{3pt}
\begin{tabular}{lrrrrr}
\toprule
\textbf{Dataset} & \textbf{\#Users} & \textbf{\#POIs} & \textbf{\#Visits} & \textbf{Coverage} & \textbf{Density} \\
\midrule
Yelp-2018  & 31,668 & 38,048 & 1,561,406 & native & 0.13\% \\
FSQ-NYC    &  1,024 &  3,298 &    27,149 & 97.4\% & 0.80\% \\
FSQ-TKY    &  2,293 &  7,098 &    57,307 & 96.1\% & 0.35\% \\
\bottomrule
\end{tabular}
\caption{Dataset statistics after filtering. Foursquare coverage is measured after cross-source augmentation.}
\label{tab:data}
\end{table}

\begin{table*}[!t]
\centering
\small
\setlength{\tabcolsep}{2pt}
\renewcommand{\arraystretch}{1.1}
\begin{tabular}{llcccccccccccc}
\toprule
& & \multicolumn{4}{c}{\textbf{Yelp-2018}}
  & \multicolumn{4}{c}{\textbf{FSQ-NYC}}
  & \multicolumn{4}{c}{\textbf{FSQ-TKY}} \\
\cmidrule(lr){3-6}\cmidrule(lr){7-10}\cmidrule(lr){11-14}
\textbf{Grp} & \textbf{Method}
  & R@10 & R@20 & N@10 & N@20
  & R@10 & R@20 & N@10 & N@20
  & R@10 & R@20 & N@10 & N@20 \\
\midrule
\multirow{2}{*}{G1}
  & BPR-MF    & 0.0421 & 0.0683 & 0.0258 & 0.0315 & 0.0512 & 0.0834 & 0.0315 & 0.0387 & 0.0487 & 0.0793 & 0.0298 & 0.0365 \\
  & NGCF      & 0.0534 & 0.0872 & 0.0329 & 0.0402 & 0.0647 & 0.1053 & 0.0402 & 0.0494 & 0.0615 & 0.1002 & 0.0385 & 0.0472 \\
\midrule
\multirow{2}{*}{G2}
  & SGL       & 0.0589 & 0.0945 & 0.0367 & 0.0448 & 0.0708 & 0.1152 & 0.0447 & 0.0548 & 0.0674 & 0.1097 & 0.0427 & 0.0523 \\
  & SimGCL    & 0.0612 & 0.0983 & 0.0394 & 0.0481 & 0.0734 & 0.1193 & 0.0468 & 0.0573 & 0.0698 & 0.1137 & 0.0448 & 0.0548 \\
\midrule
\multirow{4}{*}{G3}
  & HCCF+Zero & 0.0643 & 0.1028 & 0.0412 & 0.0502 & 0.0761 & 0.1238 & 0.0489 & 0.0597 & 0.0723 & 0.1178 & 0.0464 & 0.0568 \\
  & HCCF+Mean & 0.0671 & 0.1074 & 0.0431 & 0.0526 & 0.0793 & 0.1284 & 0.0512 & 0.0625 & 0.0754 & 0.1226 & 0.0487 & 0.0595 \\
  & DHCF+Zero & 0.0615 & 0.0987 & 0.0398 & 0.0486 & 0.0724 & 0.1176 & 0.0462 & 0.0565 & 0.0692 & 0.1127 & 0.0443 & 0.0542 \\
  & DHCF+Mean & 0.0638 & 0.1022 & 0.0415 & 0.0507 & 0.0758 & 0.1229 & 0.0487 & 0.0593 & 0.0718 & 0.1168 & 0.0461 & 0.0564 \\
\midrule
\multirow{3}{*}{G4}
  & DHCF-Dyn  & 0.0583 & 0.0936 & 0.0371 & 0.0453 & 0.0687 & 0.1117 & 0.0438 & 0.0535 & 0.0652 & 0.1062 & 0.0416 & 0.0509 \\
  & ReHDM     & 0.0628 & 0.1005 & 0.0403 & 0.0492 & 0.0741 & 0.1202 & 0.0475 & 0.0580 & 0.0704 & 0.1145 & 0.0452 & 0.0553 \\
  & HIRE      & \underline{0.0784} & \underline{0.1238} & \underline{0.0511} & \underline{0.0621} & \underline{0.0901} & \underline{0.1451} & \underline{0.0590} & \underline{0.0712} & \underline{0.0859} & \underline{0.1386} & \underline{0.0563} & \underline{0.0681} \\
\midrule
\multirow{4}{*}{G5}
  & LLM-Rec   & 0.0714 & 0.1139 & 0.0462 & 0.0598 & 0.0835 & 0.1351 & 0.0542 & 0.0687 & 0.0798 & 0.1294 & 0.0517 & 0.0657 \\
  & RLMRec    & 0.0752 & 0.1192 & 0.0487 & 0.0584 & 0.0872 & 0.1408 & 0.0569 & 0.0674 & 0.0831 & 0.1347 & 0.0543 & 0.0641 \\
  & MoDiCF    & 0.0734 & 0.1165 & 0.0473 & 0.0571 & 0.0851 & 0.1378 & 0.0554 & 0.0662 & 0.0812 & 0.1318 & 0.0529 & 0.0629 \\
  & DGMRec    & 0.0771 & 0.1221 & 0.0501 & 0.0609 & 0.0884 & 0.1427 & 0.0579 & 0.0697 & 0.0843 & 0.1364 & 0.0553 & 0.0668 \\
\midrule
\multirow{1}{*}{Ours}
  & \textbf{\method}
  & \textbf{0.0831} & \textbf{0.1314} & \textbf{0.0548} & \textbf{0.0667}
  & \textbf{0.0963} & \textbf{0.1552} & \textbf{0.0638} & \textbf{0.0779}
  & \textbf{0.0914} & \textbf{0.1483} & \textbf{0.0601} & \textbf{0.0734} \\
\bottomrule
\end{tabular}
\caption{Performance under random modality masking at $\rho_{\text{miss}}=0.6$. Values are means over five seeds; bold and underlining denote the highest overall and baseline means, respectively. Detailed results including standard deviations are reported in Supplementary Section~C}
\label{tab:main}
\end{table*}

\subsubsection{Evaluation Protocol.}
We evaluate chronological next POI ranking with Recall@10/20 and NDCG@10/20. Each held-out target is ranked against the same full catalog excluding POIs in the user's training history. The graph, prompts, node representations, retrieval pool, and incidences are cached at the training cutoff and are not updated by validation or test interactions. All methods share the split, mask, and candidate set for each seed.

\subsubsection{Baselines.}
Table~\ref{tab:main} groups 15 baselines into five families: G1, classic collaborative filtering and bipartite graph methods, BPR-MF~\cite{rendle2012bpr} and NGCF~\cite{wang2019neural}; G2, graph contrastive methods, SGL~\cite{wu2021self} and SimGCL~\cite{yu2022graph}; G3, static hypergraph methods, HCCF~\cite{xia2022hypergraph} and DHCF~\cite{ji2020dual}, each with zero and mean imputation; G4, DHCF-Dyn, our dynamic-topology extension of DHCF following Wei et al.~\cite{wei2022dynamic}, the next POI model ReHDM~\cite{li2025beyond}, and HIRE~\cite{lin2025unified}, which jointly completes missing modalities while constructing and sparsifying a heterogeneous hypergraph; and G5, semantic and multimodal methods, LLM-Rec~\cite{lyu2024llm}, RLMRec~\cite{ren2024representation}, MoDiCF~\cite{li2025generating}, which uses diffusion and counterfactual learning, and DGMRec~\cite{kim2025disentangling}, which separates shared and modality-specific factors. We reproduce official implementations, and select configurations by the same validation NDCG@20. Baseline reproduction details are provided in Supplementary Section~D.

\subsubsection{Implementation Details.}
HyperAgent4POI uses two DSMP layers and a frozen Llama-3-8B backbone with three role-specific LoRA adapters. Each adapter has rank $r=16$, scaling $\alpha=32$, and is applied to the query and value projections. Local Expert and Semantic Aggregator share the aggregation adapter but receive different inputs. POI reviews are concatenated chronologically and truncated to 64 Llama tokens. For POIs with multiple images, we average at most five frozen CLIP ViT-B/32 embeddings~\cite{radford2021learning}; text features use frozen \texttt{all-MiniLM-L6-v2} embeddings~\cite{wang2020minilm}. Their 512- and 384-dimensional features are mapped to eight continuous prefix embeddings. FAISS retrieves candidates by cosine similarity over projected POI states. Retrieval is recomputed at every layer over the bounded candidate pool and is excluded from gradient propagation.

We train for up to 50 epochs with Adam, learning rate $10^{-4}$, weight decay $10^{-5}$, recommendation batch size 512, and early stopping on validation NDCG@20 with patience 10. Unless otherwise stated, $\tau=0.75$, $K_{\mathrm{ret}}=5$, $\gamma=0.1$, $\lambda=0.5$, $T=0.07$, and $\varepsilon_{\mathrm{num}}=\varepsilon_{\mathrm{lab}}=10^{-6}$. Agent sequences are micro-batched by role and length in groups of 32. All experiments are conducted on NVIDIA A100 80\,GB GPUs using five random seeds. Sensitivity results for $\tau$, $K_{\mathrm{ret}}$, $\gamma$, and LoRA rank are provided in Supplementary Section~E.

\subsection{RQ1: Main Performance Comparison}
\label{sec:rq1}

At $\rho_{\text{miss}}=0.6$, Table~\ref{tab:main} shows that \method{} attains the highest mean in all 12 dataset--metric combinations. Static hypergraph baselines generally outperform G1 and G2, whereas DHCF-Dyn trails HCCF+Zero in all entries, indicating that dynamic incidence estimation remains sensitive to incomplete representations. Among methods designed for incomplete multimodal recommendation, HIRE is the strongest baseline in every entry. \method{} exceeds HIRE in NDCG@20 by 7.4\%, 9.4\%, and 7.8\% on Yelp-2018, FSQ-NYC, and FSQ-TKY, respectively. Across five matched random seeds, the standard deviations of NDCG@20 for \method{}, HIRE, and DGMRec do not exceed 0.0023 on any dataset. Moreover, the improvements of \method{} over all baselines remain statistically significant after Holm--Bonferroni correction
($p_{\mathrm{Holm}}<0.05$).

\subsection{RQ2: Robustness to Modality Missingness}
\label{sec:rq2}

Table~\ref{tab:robustness} reports the complete NDCG@20 trajectories for HIRE and \method{}. Additional method comparisons and structured-missingness results are provided in Supplementary Tables~C1 and~F1, respectively. \method{} attains the highest NDCG@20 at each evaluated rate. As $\rho_{\text{miss}}$ increases from 0.2 to 0.8, NDCG@20 decreases by 25.6--26.7\% for HIRE and by 17.3--18.2\% for \method{}. At $\rho_{\text{miss}}=0.8$, the performance gap widens, while \method{} shows a smaller relative decline. This result suggests that layer-wise contextual completion becomes increasingly beneficial when less modality information is directly available.

\begin{table}[t]
\centering
\small
\setlength{\tabcolsep}{2.2pt}
\begin{tabular}{llcccc}
\toprule
\textbf{Dataset} & \textbf{Method} & $\rho=.2$ & $\rho=.4$ & $\rho=.6$ & $\rho=.8$ \\
\midrule
\multirow{2}{*}{Yelp} & HIRE & .0718 & .0671 & .0621 & .0534 \\
 & \textbf{\method} & \textbf{.0762} & \textbf{.0718} & \textbf{.0667} & \textbf{.0630} \\
\multirow{2}{*}{NYC} & HIRE & .0831 & .0775 & .0712 & .0609 \\
 & \textbf{\method} & \textbf{.0897} & \textbf{.0841} & \textbf{.0779} & \textbf{.0734} \\
\multirow{2}{*}{TKY} & HIRE & .0793 & .0739 & .0681 & .0581 \\
 & \textbf{\method} & \textbf{.0839} & \textbf{.0793} & \textbf{.0734} & \textbf{.0692} \\
\bottomrule
\end{tabular}
\caption{Mean NDCG@20 across modality-missing rates over five seeds.}
\label{tab:robustness}
\end{table}

\begin{figure}[!t]
\centering
\includegraphics[width=0.8\linewidth]{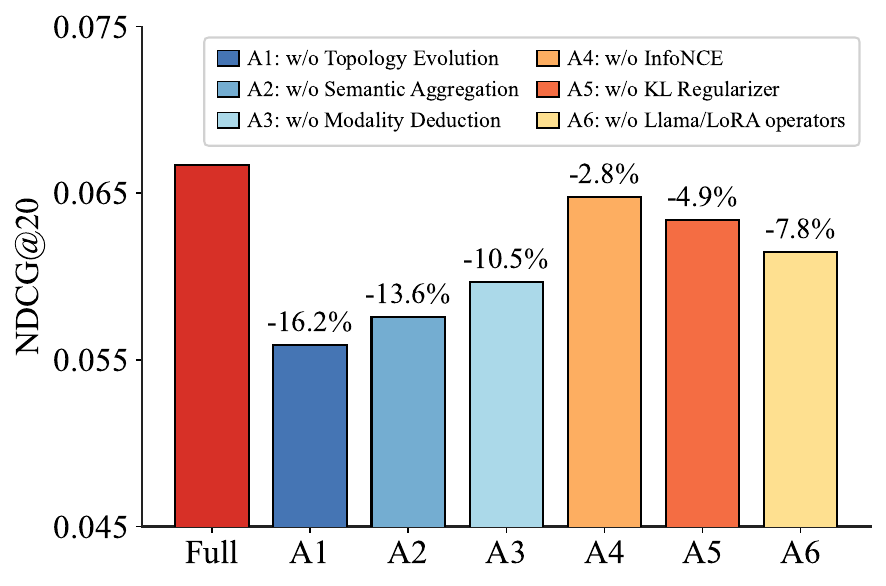}
\caption{Ablation on Yelp-2018 ($\rho_{\text{miss}}=0.6$) over five seeds; labels denote NDCG@20 reductions from the full model.}
\label{fig:ablation}
\end{figure}

\subsection{RQ3: Ablation Study}
\label{sec:ablation}

Figure~\ref{fig:ablation} reports six Yelp-2018 ablations at $\rho_{\text{miss}}=0.6$: A1 fixes the original topology; A2 replaces semantic aggregation with mean pooling; A3 removes Modality Deduction; A4--A5 remove InfoNCE and topology KL; and A6 replaces the role-conditioned Llama/LoRA operators with parameter-matched role-specific residual MLPs while preserving inputs, candidate pools, objectives, training, and readout. Implementation details for A6 are provided in Supplementary Section~F--H. The NDCG@20 reductions for A1--A6 are 16.2\%, 13.6\%, 10.5\%, 2.8\%, 4.9\%, and 7.8\%, respectively; A6 obtains 0.0615. A1--A3 demonstrate the importance of the three layer operations, while A6 shows that the improvement is not solely due to increased trainable parameters, as the Llama/LoRA operators outperform parameter-matched MLPs.

\subsection{RQ4: Layer-wise Completion}
\label{sec:mechanism}

To isolate the effect of completion frequency, we compare \emph{NoComp} (learned missing-modality embeddings), \emph{PreComp} (one deduction before propagation), and \emph{Layerwise DSMP} (one deduction in each layer) under a matched backbone, supervision, and computational budget. The diagnostics comprise mean NDCG@20 across three datasets, completion cosine similarity (CompCos) on supervised Yelp-2018 masks, and
\begin{equation}
\mathrm{RepDisc}=\frac{1}{|\mathcal P_M|}\sum_{p\in\mathcal P_M}\left[1-\cos\!\left(\mathbf x_{p,\mathrm{mask}}^{(L)},\mathbf x_{p,\mathrm{full}}^{(L)}\right)\right],
\label{eq:repdisc}
\end{equation}
Here, $\mathcal P_M$ is the set of POIs with supervised masks, while $\mathbf x_{p,\mathrm{mask}}^{(L)}$ and $\mathbf x_{p,\mathrm{full}}^{(L)}$ are the final states obtained from masked and restored inputs, respectively, under identical model parameters. The restored inputs are used only for this analysis.

\begin{table}[t]
\centering
\small
\setlength{\tabcolsep}{3pt}
\begin{tabular}{lccc}
\toprule
\textbf{Method} & \textbf{Avg.\ N@20} & \textbf{CompCos} $\uparrow$ & \textbf{RepDisc} $\downarrow$ \\
\midrule
NoComp         & 0.0640 & --    & 0.631 \\
PreComp        & 0.0696 & 0.712 & 0.387 \\
Layerwise DSMP & 0.0727 & 0.784 & 0.241 \\
\bottomrule
\end{tabular}
\caption{Completion diagnostics. Avg.\ N@20 is averaged across three datasets; the remaining metrics are computed post hoc using supervised Yelp-2018 masks.}
\label{tab:mechanism}
\end{table}

Layerwise DSMP improves mean NDCG@20 and CompCos over PreComp by $4.5\%$ and $10.1\%$, respectively, while reducing RepDisc by $37.7\%$ (0.241 vs.\ 0.387). The joint improvement in ranking and representation quality indicates that revisiting completion after each hyperedge-context update preserves more task-relevant information than a fixed one-shot reconstruction.
\subsection{RQ5: Offline and Online Efficiency}
\label{sec:efficiency}

Across five Yelp-2018 runs, training requires 64 GPU-hours. On four A100 GPUs, a soft incidence refresh takes 22 minutes, reaches 28\,GB peak memory, and processes 3.76M agent sequences, corresponding to 98.8 sequences and 34.7\,ms per POI. Cached serving takes 14.7\,ms per 1,024 queries ($+2.3$\,ms vs. HGCN; $-16.3$\,ms vs. DHCF-Dyn). DSMP has 23.6M trainable parameters (0.29\% of the backbone). Further cost details are provided in Supplementary Section~H. Thus, semantic computation occurs during refresh, while serving uses representations computed with soft incidences and dot-product readout.


\section{Conclusion}
\label{sec:conclusion}

We introduced HyperAgent4POI for next POI recommendation with incomplete multimodal POI content. DSMP jointly performs semantic hyperedge aggregation, soft incidence refinement, and modality completion within each propagation layer. Experiments on three LBSN benchmarks demonstrate consistent ranking improvements across modality-missing rates, while ablations verify the contributions of the DSMP operations and role-conditioned semantic operators. Because LLM computation is confined to training and offline representation refresh, cached node representations support efficient online ranking without LLM calls.

\bibliography{aaai2027}


\end{document}